\documentclass[twocolumn,showpacs,preprintnumbers,amsmath,amssymb,prl,superscriptaddress]{revtex4}

\usepackage{graphicx} 
\usepackage{dcolumn}  
\usepackage{colordvi}

\graphicspath{{ps}}

\newcommand{\dz}{\ensuremath{D^0}}
\newcommand{\dzbar}{\ensuremath{\overline{D}{}^0}}
\newcommand{\mev}{\ensuremath{\mathrm{MeV}}}
\newcommand{\gev}{\ensuremath{\mathrm{GeV}}}

\begin{document}

\preprint{\vbox{ \hbox{   }
    \hbox{}
    \hbox{}
    \hbox{} 
}}

\title { \quad\\[0.5cm]  Evidence for \dz--\dzbar\ Mixing}

\affiliation{Budker Institute of Nuclear Physics, Novosibirsk}
\affiliation{University of Cincinnati, Cincinnati, Ohio 45221}
\affiliation{Justus-Liebig-Universit\"at Gie\ss{}en, Gie\ss{}en}
\affiliation{The Graduate University for Advanced Studies, Hayama}
\affiliation{Gyeongsang National University, Chinju}
\affiliation{Hanyang University, Seoul}
\affiliation{University of Hawaii, Honolulu, Hawaii 96822}
\affiliation{High Energy Accelerator Research Organization (KEK), Tsukuba}
\affiliation{Hiroshima Institute of Technology, Hiroshima}
\affiliation{University of Illinois at Urbana-Champaign, Urbana, Illinois 61801}
\affiliation{Institute of High Energy Physics, Chinese Academy of Sciences, Beijing}
\affiliation{Institute of High Energy Physics, Vienna}
\affiliation{Institute of High Energy Physics, Protvino}
\affiliation{Institute for Theoretical and Experimental Physics, Moscow}
\affiliation{J. Stefan Institute, Ljubljana}
\affiliation{Kanagawa University, Yokohama}
\affiliation{Korea University, Seoul}
\affiliation{Swiss Federal Institute of Technology of Lausanne, EPFL, Lausanne}
\affiliation{University of Ljubljana, Ljubljana}
\affiliation{University of Maribor, Maribor}
\affiliation{University of Melbourne, School of Physics, Victoria 3010}
\affiliation{Nagoya University, Nagoya}
\affiliation{Nara Women's University, Nara}
\affiliation{National Central University, Chung-li}
\affiliation{National United University, Miao Li}
\affiliation{Department of Physics, National Taiwan University, Taipei}
\affiliation{H. Niewodniczanski Institute of Nuclear Physics, Krakow}
\affiliation{Nippon Dental University, Niigata}
\affiliation{Niigata University, Niigata}
\affiliation{University of Nova Gorica, Nova Gorica}
\affiliation{Osaka City University, Osaka}
\affiliation{Osaka University, Osaka}
\affiliation{Panjab University, Chandigarh}
\affiliation{Peking University, Beijing}
\affiliation{Princeton University, Princeton, New Jersey 08544}
\affiliation{RIKEN BNL Research Center, Upton, New York 11973}
\affiliation{Saga University, Saga}
\affiliation{University of Science and Technology of China, Hefei}
\affiliation{Seoul National University, Seoul}
\affiliation{Shinshu University, Nagano}
\affiliation{Sungkyunkwan University, Suwon}
\affiliation{University of Sydney, Sydney, New South Wales}
\affiliation{Tata Institute of Fundamental Research, Mumbai}
\affiliation{Toho University, Funabashi}
\affiliation{Tohoku Gakuin University, Tagajo}
\affiliation{Tohoku University, Sendai}
\affiliation{Department of Physics, University of Tokyo, Tokyo}
\affiliation{Tokyo Institute of Technology, Tokyo}
\affiliation{Tokyo Metropolitan University, Tokyo}
\affiliation{Tokyo University of Agriculture and Technology, Tokyo}
\affiliation{Virginia Polytechnic Institute and State University, Blacksburg, Virginia 24061}
\affiliation{Yonsei University, Seoul}
  \author{M.~Stari\v c}\affiliation{J. Stefan Institute, Ljubljana} 
  \author{B.~Golob}\affiliation{University of Ljubljana, Ljubljana}\affiliation{J. Stefan Institute, Ljubljana} 
  \author{K.~Abe}\affiliation{High Energy Accelerator Research Organization (KEK), Tsukuba} 
  \author{K.~Abe}\affiliation{Tohoku Gakuin University, Tagajo} 
  \author{I.~Adachi}\affiliation{High Energy Accelerator Research Organization (KEK), Tsukuba} 
  \author{H.~Aihara}\affiliation{Department of Physics, University of Tokyo, Tokyo} 
  \author{K.~Arinstein}\affiliation{Budker Institute of Nuclear Physics, Novosibirsk} 
  \author{T.~Aushev}\affiliation{Swiss Federal Institute of Technology of Lausanne, EPFL, Lausanne}\affiliation{Institute for Theoretical and Experimental Physics, Moscow} 
  \author{T.~Aziz}\affiliation{Tata Institute of Fundamental Research, Mumbai} 
  \author{S.~Bahinipati}\affiliation{University of Cincinnati, Cincinnati, Ohio 45221} 
% ...
\collaboration{The Belle Collaboration}

\date{\today}

\begin{abstract}
  We observe evidence for \dz--\dzbar\ mixing by measuring the
  difference in apparent lifetime when a \dz\ meson decays to the $CP$
  eigenstates $K^+K^-$ and $\pi^+\pi^-$, and when it decays to the final state
  $K^-\pi^+$. We find
  $y_{CP}=(1.31\pm 0.32({\rm stat.})\pm 0.25({\rm syst.}))\%$, 
  3.2 standard deviations from zero.
  We also search for a $CP$ asymmetry between \dz\ and \dzbar\ decays;
  no evidence for $CP$ violation is found.
  These results are based on 540~fb$^{-1}$ of data recorded
  by the Belle detector at the KEKB $e^+e^-$ collider.
\end{abstract}
\maketitle

The phenomenon of mixing between a particle and its anti-particle has
been observed in several systems of neutral mesons~\cite{PDG,BsMix}:
neutral kaons, $B_d^0$, and most recently $B_s^0$ mesons.
In this paper we present
evidence for \dz--\dzbar\ mixing~\cite{babar_kpi_new}.

The time evolution of a \dz or \dzbar\ is governed by the mixing
parameters $x=(M_1-M_2)/\Gamma$ and $y=(\Gamma_1-\Gamma_2)/2\Gamma$,
where $M_{1,2}$ and $\Gamma_{1,2}$ are the 
masses and widths, respectively, of the mass eigenstates, and $\Gamma
=(\Gamma_1+\Gamma_2)/2$. For no mixing, $x=y=0$.
Within the Standard Model (SM) the rate of $D$-mixing is expected to
be small due to the near degeneracy of the $s$ and $d$ quark masses
relative to the $W$ mass, and the small value of the $b$ quark
couplings. 
Predictions for $x$ and $y$ are dominated by 
non-perturbative processes that are difficult to calculate \cite{bigi-uraltsev,falk}. 
The largest predictions are 
$|x|,~|y| \sim {\cal{O}}(10^{-2})$ \cite{falk}. 
Loop diagrams including new, as-yet-unobserved particles could
significantly affect the experimental values~\cite{petrov}.
$CP$-violating effects in $D$-mixing would be a clear signal of new physics,
as $CP$ violation ($CPV$) is expected to be very small in the SM~\cite{bigi-sanda}.

Both semileptonic and hadronic $D$ decays have been used
to constrain $x$ and $y$ \cite{PDG}.
Here we study the decays to $CP$ eigenstates $D^0\to K^+K^-$ and
$D^0\to\pi^+\pi^-$; 
treating the decay time distributions as exponential, we measure 
the quantity 
\begin{equation}
  y_{CP}=\frac{\tau(K^-\pi^+)}{\tau(K^+K^-)}-1,
\end{equation}
where $\tau(K^+K^-)$ and $\tau(K^-\pi^+)$ are the lifetimes of
$D^0\to K^+K^-$ (or $\pi^+\pi^-$) and $D^0\to K^-\pi^+$ decays \cite{conjugate}. It can 
be shown that $y_{CP}=y\cos{\phi}-\frac{1}{2}A_Mx\sin{\phi}$~\cite{Bergman}, 
where $A_M$ parameterizes $CPV$ in mixing and $\phi$ is a weak phase. 
If $CP$ is conserved, $A_M=\phi=0$ and $y_{CP}=y$. To date several 
measurements of $y_{CP}$ have been reported~\cite{ycp_exp}; the
average value is $\sim$2 standard deviations ($\sigma$) above zero. Our measurement yields 
a nonzero value of $y_{CP}$ with $>3\sigma$
significance. We also search for $CPV$ by measuring the quantity
\begin{equation}
  A_{\Gamma}=\frac{\tau(\dzbar\to K^-K^+)-\tau(D^0\to K^+K^-)}
    {\tau(\dzbar\to K^-K^+)+\tau(D^0\to K^+K^-)};
\end{equation}
this observable equals $A_\Gamma=\frac{1}{2}A_My\cos{\phi}-x\sin{\phi}$~\cite{Bergman}.

Our results are based on 540~fb$^{-1}$ of data recorded
by the Belle experiment~\cite{Belle} at the KEKB asymmetric-energy
$e^+e^-$~collider \cite{KEKB}, running at the center-of-mass (CM) energy of
the $\Upsilon(4S)$ resonance and 60~MeV below. 
To avoid bias, details of the analysis procedure were finalized without 
consulting quantities sensitive to $y_{CP}$ and $A_\Gamma$.

\begin{figure}[t]
  \includegraphics[width=8.57cm]{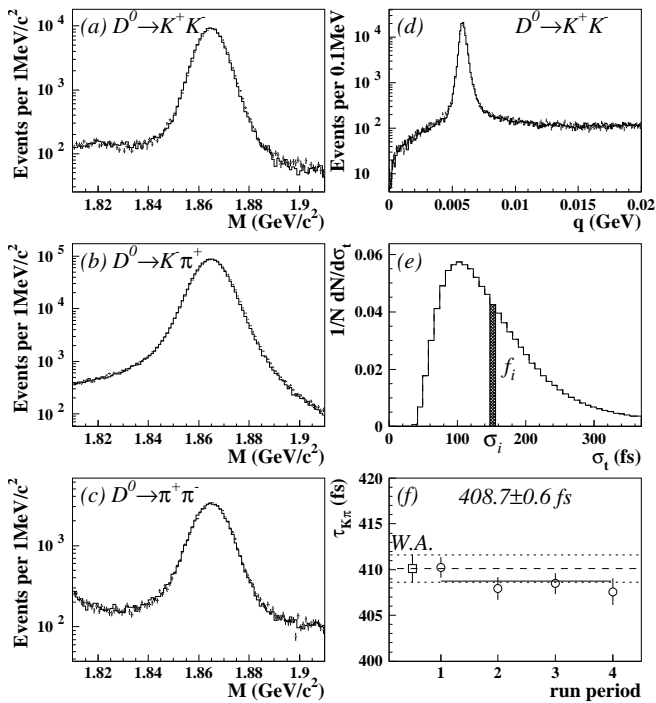}
  \caption{$M$ distribution of selected events (with
    $|\Delta q| < 0.80\,\mev$ and $\sigma_t<370$~fs)
    for (a) $K^+K^-$, 
    (b) $K^-\pi^+$ and (c) $\pi^+\pi^-$ final states. The histogram shows
    the tuned MC distribution. (d) $q$ distribution (with $|\Delta M|/\sigma_M<2.3$
    and $\sigma_t<370$~fs) for the $K^+K^-$ final state. 
    (e) Normalized distribution of errors $\sigma_t$ on the decay time $t$
    for $D^0\to K^-\pi^+$, showing the construction of the resolution function
    using the fraction $f_i$ in the bin with $\sigma_t = \sigma_i$.
    (f) Fitted lifetime of $D^0$ mesons in the $K^-\pi^+$ final state
    in four running periods with slightly different conditions, and  
    the result of a fit to a constant. The world average value (W.A.) is also shown.}
  \label{fig1}
\end{figure}

The Belle detector is described in detail elsewhere~\cite{Belle}:
it includes in particular a silicon vertex detector ~\cite{SVD2}, a 
central drift chamber,
an array of aerogel Cherenkov counters,
and time-of-flight scintillation counters.
We reconstruct $D^{\ast +}\to D^0\pi^+_s$ decays with a characteristic
slow pion $\pi_s$, and $D^0\to K^+K^-$, $K^-\pi^+$, and $\pi^+\pi^-$.
Each track is required
to have at least two associated vertex detector hits in each of the two
measuring coordinates.
To select pion and kaon candidates, we impose
standard particle identification criteria~\cite{PID}.
$D^0$ daughter tracks are refitted to 
a common vertex, and the $D^0$ production vertex is
found by constraining its momentum vector and the $\pi_s$ track to originate
from the $e^+e^-$ interaction region; confidence levels exceeding $10^{-3}$
are required for both fits.  A $D^\ast$ momentum greater than 
$2.5\,\gev/c$ (in the CM) is required to reject $D$-mesons
produced in $B$-meson decays and to suppress combinatorial background.
The proper  decay time of the
$D^0$ candidate is then calculated from the projection of the vector
joining the two vertices, $\vec{L}$, onto the $D^0$ momentum vector, 
$t=m_{\dz}\vec{L}\cdot\vec{p}/p^2$, where
$m_{\dz}$ is the nominal $D^0$ mass. The decay time uncertainty 
$\sigma_t$ is evaluated event-by-event from the covariance
matrices of the production and decay vertices.

Candidate $D^0$ mesons are selected using two kinematic observables:
the invariant mass of the $D^0$ decay products, $M$, and the energy
released in the $D^{\ast +}$ decay,
$q=(M_{D^\ast}-M-m_\pi)c^2$. $M_{D^\ast}$ is the invariant mass of
the $D^0 \pi_s$ combination and $m_\pi$ is the $\pi^+$ mass.

According to Monte Carlo (MC) simulated distributions of $t$, $M$, and $q$, 
background events fall into four 
categories: (1) combinatorial, with zero apparent lifetime; 
(2) true $D^0$ mesons combined with random slow pions (this has the
same apparent lifetime as the signal);  
(3) $D^0$ decays to three or more particles, 
and (4) other charm hadron decays. The apparent lifetime of the latter two categories is
10--30\% larger than $\tau_{D^0}$. 
Since we find differences in $M$ and
$q$ distributions between MC and data events, we perform fits to 
data distributions to obtain scaling factors for the
background fractions and signal widths, and then tune
the signal fractions and shapes in the MC  
event-by-event. 

The sample of events for the lifetime measurements
is selected using $|\Delta M|/\sigma_M$, where $\Delta M \equiv M-m_{\dz}$;
$|\Delta q| \equiv q - (m_{D^{\ast +}} - m_{D^0} - m_\pi)c^2$;
and $\sigma_t$.
The invariant mass resolution $\sigma_M$ 
varies from 5.5--6.8~MeV/$c^2$,
depending on the decay channel.
Selection criteria are chosen to minimize the expected statistical error
on $y_{CP}$, using the tuned MC:
we require $|\Delta M|/\sigma_M<2.3$, $|\Delta q| < 0.80\,\mev$, and
$\sigma_t<370$~fs. The data distributions and agreement with the tuned MC are shown 
in Figs~\ref{fig1}(a)--(d).
We find $111\times 10^3$ $K^+K^-$, $1.22\times 10^6$ $K^-\pi^+$,
and $49\times 10^3$ $\pi^+\pi^-$ signal events, 
with purities of 98\%, 99\%, and 92\% respectively. 

The relative lifetime difference $y_{CP}$ is determined from 
$D^0\to K^+K^-$, $K^-\pi^+$, and $\pi^+\pi^-$ decay time distributions
by performing a simultaneous binned maximum likelihood fit
to the three samples. Each distribution
is assumed to be a sum of signal and background contributions, with the signal
contribution being a convolution of an exponential 
and a detector resolution function,
\begin{equation}
  dN/dt = \frac{N_{\rm{sig}}}{\tau}\int e^{-t'/\tau} \cdot R(t-t')\, \mathrm{d}t' + B(t).  
\end{equation}
The resolution function $R(t-t')$ is constructed from the normalized distribution of the
decay time uncertainties $\sigma_t$ (see Fig.~\ref{fig1}(e)).
The $\sigma_t$ of a reconstructed event
ideally represents an uncertainty with a Gaussian probability density:
in this case, we take bin $i$ in the $\sigma_t$ distribution to correspond to
a Gaussian resolution term of width $\sigma_i$,
with a weight given by the fraction $f_i$ of events in that bin.
However, the distribution of ``pulls'', i.e.  
the normalized residuals $(t_{\text{rec}}-t_{\text{gen}})/\sigma_t$
(where $t_{\text{rec}}$ and $t_{\text{gen}}$ are reconstructed and generated
MC decay times), is not well-described by a Gaussian.
We find that this distribution can be fitted with a sum of three Gaussians
of different widths $\sigma_k^{\rm pull}$ and fractions $w_k$, constrained to the same mean.
We therefore choose a parameterization
\begin{equation}
  R(t-t')=\sum_{i=1}^n{f_i}\sum_{k=1}^3{w_kG(t-t';\sigma_{ik},t_0)},
  \label{resofun.eq}
\end{equation}
with $\sigma_{ik}=s_k\sigma_k^{\rm pull} \sigma_i$, where the $s_k$ are three scale
factors introduced to account for differences between the
simulated and real $\sigma_k^{\rm pull}$, and $t_0$ allows for a (common) offset
of the Gaussian terms from zero.

The background $B(t)$ is parameterized assuming two lifetime components:
an exponential and a $\delta$ function, each convolved with
corresponding resolution functions as parameterized 
by Eq.~(\ref{resofun.eq}). Separate $B(t)$ parameters for each final state are
determined by fits to the $t$ distributions of events in $M$ sidebands.
The tuned MC is used to select the sideband region 
that best reproduces the timing distribution of background events
in the signal region. 
We find good agreement between the tuned MC and data sidebands, 
with a normalized $\chi^2$ of 0.85, 0.83 and 0.83 for $KK$, $K\pi$, and $\pi\pi$ respectively. 

\begin{figure}
  \includegraphics[width=8.57cm]{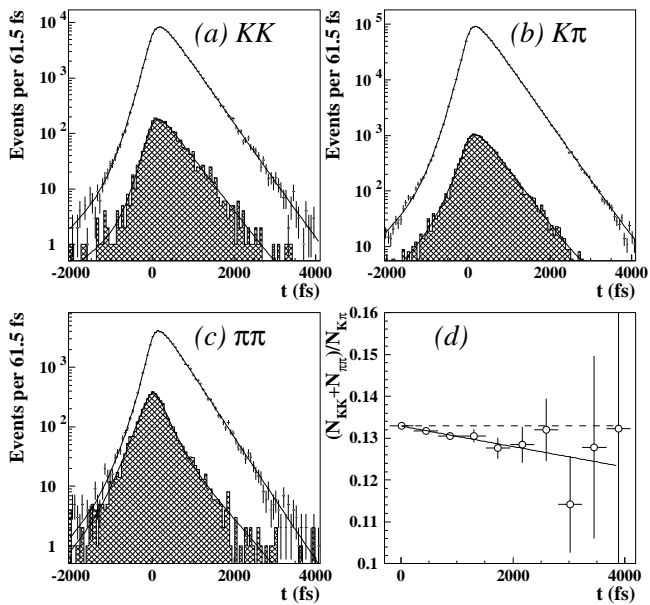}
  \caption{Results of the simultaneous fit to decay time distributions
    of (a) $D^0\to K^+K^-$, (b) $D^0\to K^-\pi^+$ and 
    (c) $D^0\to \pi^+\pi^-$ decays. The
    cross-hatched area represents background contributions, the shape of
    which was fitted using $M$ sideband events.
    (d) Ratio of decay time distributions between $D^0\to K^+K^-,
    \pi^+\pi^-$ and $D^0\to K^-\pi^+$ decays. The solid line is a fit to the data
    points.}
  \label{fig3}
\end{figure} 

The $R(t-t')$ and background parameterizations are
validated using MC and the large $D^0\to K^-\pi^+$ sample selected from data.
In the simulation, the ratio of scale factors $s_k$ 
($k=1,2,3$) is consistent between decay modes,
within small statistical uncertainties.
The offset $t_0$ is also
independent of the final state, but it changes slightly for simulated
samples describing different running periods.
Four such periods, coinciding with changes to the detector, have been
identified 
based on small variations of the mean $t$ value
for $\dz\to K^-\pi^+$ in the data.  
We perform a separate fit 
to each period and average the results to obtain the final value 
of $y_{CP}$. The free parameters of each simultaneous fit are:
$\tau_{D^0}$, $y_{CP}$, three factors $s_k$
for the $K^-\pi^+$ mode and two terms that rescale the $s_k$
in the $K^+K^-$ and $\pi^+\pi^-$ channels,
the offset $t_0$, and normalization
terms for the three decay modes.
Fits to the $D^0\to K^-\pi^+$ sample show good agreement
with the parameters of $R(t-t')$ obtained from simulation.

For the second running period, we modify Eq.~(\ref{resofun.eq}) to add
mode-dependent offsets $\Delta t$ between the first two Gaussian terms, 
making the resolution function asymmetric;
these three parameters are also left free in the fit. 
We find that such a function is required to yield the 
$\dz\to K^-\pi^+$ lifetime consistent with that in the other running periods.
(This behaviour has been reproduced with a MC model
including a small relative misalignment of the vertex detector and the
drift chamber.)
The lifetime fit results are shown in Fig.~\ref{fig1}(f):
the mean, 
$\tau_{D^0}=(408.7\pm 0.6\;\text{(stat.)})$~fs,
is in good agreement with the current world average,
$(410.1\pm 1.5)$~fs~\cite{PDG}.   

Fits to the $D^0\to K^+K^-$, $K^-\pi^+$ and
$\pi^+\pi^-$ data for the four running periods are shown 
in Fig.~\ref{fig3}(a)-(c), by summing both the data points and the fit 
functions. Averaging the fit results, we find 
$y_{CP}=(1.31\pm 0.32\;\text{(stat.)})\%$,
4.1 standard deviations from  zero. 
The agreement between the data and the fit functions is good:
$\chi^2/n_{dof}=1.08$ for $n_{dof}=289$ degrees of freedom.
Fitting
$K^+K^-/K^-\pi^+$ and $\pi^+\pi^-/K^-\pi^+$ events separately we obtain
$y_{CP}=(1.25\pm 0.39\;\text{(stat.)})\%$ and $y_{CP}=(1.44\pm 0.57\;\text{(stat.)})\%$ respectively,
in agreement with each other.
The $y_{CP}$ values for the four running periods are also consistent, with
$\chi^2/n_{dof}=1.53/3$.

To measure the $CPV$ parameter $A_\Gamma$ we separately 
determine the apparent
lifetimes of $D^0$ and $\dzbar$ in decays 
to the $CP$ eigenstates; the data is fit in four running 
periods as for $y_{CP}$. To ensure convergence of the fits, despite
the much smaller event sample, the scale factor 
for the widest Gaussian $s_3$ 
is fixed to the value obtained from the
$y_{CP}$ fit in each case.
We obtain $A_\Gamma=(0.01\pm 0.30\;\text{(stat.)})\%$, 
consistent with zero;
the quality of the fit is good, with $\chi^2/n_{dof}=1.00$ for $n_{dof}=390$.
Separate fits to the two $CP$ eigenstates find compatible values:
$A_\Gamma=(0.15\pm 0.35\;\text{(stat.)})\%$ for $K^+K^-$ and
$-(0.28\pm 0.52\;\text{(stat.)})\%$ for $\pi^+\pi^-$.

The behaviour of the fits has been tested in various ways using MC simulation.
Fits to signal events simulated with $y_{CP}=0$ reproduce this value 
(and the generated $\tau_{\dz}$) even for a sample much larger than 
the data, with ($\chi^2/n_{dof},n_{dof}) = (1.11,285)$.
Using samples of the same size as the data,
with background included, we find a satisfactory fit, 
$(\chi^2/n_{dof},n_{dof})=(1.18,289)$,
with a statistical uncertainty in agreement with the error from
the fit to the data. 
Results obtained on reweighted MC
samples that cover a wide range of $y_{CP}$ values agree with the
input within $\pm 0.04\%$.

The effect of the resolution function on the measured $y_{CP}$ 
has been tested by replacing the parameterization in Eq.~(\ref{resofun.eq})
with a single Gaussian. This describes the data poorly 
and leads to a 3.9\% shift in the fitted $\tau_{\dz}$
for a simulated $D^0\to K^-\pi^+$ sample; however, the corresponding shift
in $y_{CP}$ is only 0.01\%.
This shows that the $y_{CP}$ value returned by the fit is robust against
imperfections in the parameterization of $R(t-t')$.

The estimated systematic uncertainties are summarized in Table~\ref{tab1}.
We test for acceptance variations with decay time
by fitting the generated decay times of reconstructed
MC events.  We find no deviation, but conservatively assign the MC statistical 
error on $y_{CP}$ ($\pm 0.12\%$) to this source.
Another contribution is due to the
choice of equal $t_0$ offsets in different decay modes:
relaxing this assumption leads to $y_{CP}$ changes of $\pm 0.14\%$. 
Variation of the \dz\ mass windows changes 
$y_{CP}$ by less than $\pm 0.04\%$.
The effect of differences between backgrounds in the signal and sideband
regions is studied by repeating the fits using
MC backgrounds from signal regions;
small shifts in the data sidebands used to determine $B(t)$
are also made. The largest resulting change in $y_{CP}$,
$\pm 0.09\%$, is quoted as the
systematic error due to the background description.
Potential correlations between apparent lifetimes and opening angle
distributions (which differ between modes) have a small effect 
on $y_{CP}$: $\pm 0.02\%$.
 
The uncertainty due to the finite number of sideband events,
$\pm 0.07\%$, is
estimated by varying bin contents according to
Poisson statistics and repeating the fits. 
Comparing alternative fits where all running periods use the symmetric
resolution function~(\ref{resofun.eq}),
and the asymmetric function presently used for the second
period, we assign an additional uncertainty of $\pm 0.01\%$.
Varying selection criteria produces observable effects only in high
statistics MC samples, in the $\sigma_t$ and $|\Delta M|/\sigma_M$ cases.
The
resulting $\pm 0.11\%$ changes in $y_{CP}$ are conservatively assigned
as systematic errors. 
Finally, varying the binning of the decay-time distribution produces a
small effect, $\pm 0.01\%$.
Adding all terms in quadrature,
we obtain a systematic uncertainty on $y_{CP}$ of $\pm 0.25\%$.
The same sources dominate for $A_\Gamma$, 
but yield a smaller total systematic uncertainty, $\pm 0.15\%$.

\newlength{\zeropointzeroone}
\settowidth{\zeropointzeroone}{\ensuremath{0.01}}

\begin{table}
  \caption{\label{tab1} Sources of the systematic
    uncertainty for $y_{CP}$ and $A_\Gamma$.}
  \begin{ruledtabular}
    \begin{tabular}{l c c} 
      \multicolumn{1}{c}{Source} & $\Delta y_{CP}~[\%]$ & $\Delta A_\Gamma~[\%]$ \\ 
      \hline
      acceptance & 0.12 & 0.07   \\ 
      equal $t_0$ & 0.14 & 0.08 \\  
      $M$ window position & 0.04 & \makebox[\zeropointzeroone][r]{$<0.01$} \\  
      signal/sideband background differences & 0.09 & 0.06 \\  
      opening angle distributions & 0.02 & - - \\  
      background distribution $B(t)$ & 0.07 & 0.07 \\ 
      (a)symmetric resolution function & 0.01 & 0.01 \\ 
      selection variation & 0.11 & 0.05 \\ 
      binning of $t$ distribution  & 0.01 & 0.01 \\ 
      \hline
      Total & 0.25  & 0.15 
    \end{tabular}
  \end{ruledtabular}
\end{table}

In summary, we measure the relative difference of the apparent lifetime of
$D^0$ mesons between decays to $CP$-even eigenstates and the $K^-\pi^+$
final state to be 
\begin{equation}
  y_{CP}=(1.31\pm 0.32({\rm stat.})\pm 0.25({\rm syst.}))\%.
\end{equation}
Combining the errors in quadrature, we find a confidence level of only
$6\times 10^{-4}$ for the $y_{CP}=0$ hypothesis.
We interpret this result as evidence for mixing in the \dz--\dzbar\
system, regardless of possible $CPV$. 
The effect is presented visually in Fig.~\ref{fig3}(d), which shows the ratio
of decay time distributions for $D^0\to  K^+K^-, \pi^+\pi^-$ and
$D^0\to K^-\pi^+$ decays. 
We also search for $CP$ violation by separately measuring decay
times of $D^0$ and $\dzbar$ mesons in $CP$-even final states.
We find an asymmetry consistent with zero,
\begin{equation}
  A_\Gamma=(0.01\pm 0.30({\rm stat.})\pm 0.15({\rm syst.}))\%.
\end{equation}

\begin{acknowledgments}
We thank the KEKB group for excellent operation of the
accelerator, the KEK cryogenics group for efficient solenoid
operations, and the KEK computer group and
the NII for valuable computing and Super-SINET network
support.  We acknowledge support from MEXT and JSPS (Japan);
ARC and DEST (Australia); NSFC and KIP of CAS (China);
DST (India); MOEHRD, KOSEF and KRF (Korea);
KBN (Poland); MES and RFAAE (Russia); ARRS (Slovenia); SNSF (Switzerland);
NSC and MOE (Taiwan); and DOE (USA).
\end{acknowledgments}


\begin{thebibliography}{99}
\bibitem{PDG} W.-M.~Yao {\it et al.} (Particle Data Group), J.\ Phys.\ G{\bf 33}, 1 (2006). 
\bibitem{BsMix} A.~Abulencia {\it et al.} (CDF Collaboration),
  Phys.\ Rev.\ Lett.\ {\bf 97}, 242003 (2006); V.M.~Abazov {\it et al.} 
  (D0 Collaboration), Phys.\ Rev.\ Lett.\ {\bf 97}, 021802 (2006).
\bibitem{babar_kpi_new} During the preparation of this paper, 
  we were made aware of a positive $D$-mixing result using different observables, by
  B.~Aubert {\it et al.} (BaBar Collaboration),
  \texttt{arXiv:hep-ex/0703020}, submitted to Phys.\ Rev.\ Lett.
\bibitem{bigi-uraltsev} I.I.~Bigi, N.~Uraltsev, Nucl.\ Phys.\ B {\bf 592}, 92 (2001). 
\bibitem{falk} A.F.~Falk {\it et al.}, Phys.\ Rev.\ D{\bf 65}, 054034 (2002);
  A.F.~Falk {\it et al.}, Phys.\ Rev.\ D{\bf 69}, 114021 (2004).
\bibitem{petrov} A.A.~Petrov, Int.\ J.\ Mod.\ Phys.\ A{\bf 21}, 5686 (2006);
  E.~Golowich, S.~Pakvasa, A.A.~Petrov, \texttt{arXiv:hep-ph/0610039}. 
\bibitem{bigi-sanda} I.I.~Bigi, A.I.~Sanda, \emph{CP violation}
  (Cambridge University Press, Cambridge, 2000), p.\ 257.
\bibitem{conjugate} Charge conjugate modes are implied unless
  explicitly stated otherwise.
\bibitem{Bergman} S. Bergmann {\it et al.}, Phys. Lett. B {\bf 486},
  418 (2000). 
\bibitem{ycp_exp} E.M.~Aitala {\it et al.} (E791 Collaboration),
  Phys.\ Rev.\ Lett.\ {\bf 83}, 32 (1999); 
  J.M.~Link {\it et al.} (Focus Collaboration), Phys.\ Lett.\ B{\bf 485}, 62
  (2000); S.E.~Csorna {\it et al.} (CLEO Collaboration), Phys.\ Rev.\ D{\bf 65}, 092001
  (2002); K.~Abe {\it et al.} (Belle Collaboration), Phys.\ Rev.\ Lett.\  
  {\bf 88}, 162001 (2002); 
  B.~Aubert {\it et al.} (BaBar Collaboration), Phys.\ Rev.\ Lett.\ {\bf 91}, 121801 (2003).
\bibitem{Belle} A.~Abashian {\it et al.} (Belle Collaboration),
  Nucl.\ Instr.\ Meth.\ A{\bf 479}, 117 (2002).
\bibitem{KEKB} S.~Kurokawa, E.~Kikutani, Nucl.\ Instr.\ Meth.\ A{\bf
  499}, 1 (2003), and other papers in this volume. 
\bibitem{SVD2} Z.~Natkaniec {\it et al.} (Belle SVD2 group), 
  Nucl.\ Instr.\ Meth.\ A{\bf 560}, 1 (2006).
\bibitem{PID} E.~Nakano, Nucl.\ Instr.\ Meth.\ A{\bf 494}, 402 (2002).


\end{thebibliography}
\end{document}